\documentclass[prl,aps,amsmath,amsfonts,showpacs,twocolumn,floatfix]{revtex4}

\usepackage{bm}
\usepackage{graphicx}
\usepackage{color}

\def\be{\begin{equation}}
\def\ee{\end{equation}}

\def\br{{\bf r}}
\def\bp{{\bf p}}
\def\bq{{\bf q}}
\def\bv{{\bf v}}

\newcommand{\corr}[1]{\langle #1\rangle}
\newcommand{\ccorr}[1]{\langle\langle #1\rangle\rangle}

\def\Re{\mathop{\rm Re}}
\def\Im{\mathop{\rm Im}}

\def\eps{\varepsilon}

\def\gcf{g_{c}}

\begin{document}

\title{Universal Broadening of the Bardeen-Cooper-Schrieffer Coherence Peak of Disordered Superconducting Films}
\author{M. V. Feigel'man}
\affiliation{L. D. Landau Institute for Theoretical Physics,
142432 Chernogolovka, Russia}
\affiliation{Moscow Institute of Physics and Technology, 141700 Moscow, Russia}

\author{M. A. Skvortsov}
\affiliation{L. D. Landau Institute for Theoretical Physics,
142432 Chernogolovka, Russia}
\affiliation{Moscow Institute of Physics and Technology, 141700 Moscow, Russia}

\date{\today}

\begin{abstract}
In disordered superconductors, the local pairing field
fluctuates in space, leading to the smearing of the BCS peak
in the density of states and appearance of the subgap tail states.
We analyze the universal mesoscopic contributions to these effects
and show that they are enhanced by the Coulomb repulsion.
In the vicinity of the quantum critical point, where superconductivity
is suppressed by the ``fermionic mechanism'', strong smearing of the
peak due to mesoscopic
fluctuations is predicted.
\end{abstract}


% PACS scheme 2010:
% 74.78.-w Superconducting films and low-dimensional structures
% 74.20.-z Theories and models of superconducting state
% 74.81.-g Inhomogeneous superconductors and superconducting systems

\maketitle

Superconductive ($s$-wave) state is characterized by a gap $\Delta$ in the
quasiparticle spectrum and the coherence peak (CP) in the density of
states (DOS) above the gap,
$\rho(E) = \rho_0 \Re E/\sqrt{E^2-\Delta^2}$.
According to classical results~\cite{AG1958,Anderson1959},
impurity scattering does not affect this picture,
as long as the time-reversal invariance (TRI) is not broken.
Yet, a number of experiments demonstrate considerable\- suppression
of the CP and appearance of subgap ($E < \Delta$) states with
the increase of disorder~\cite{Dynes,Tashiro2008,ONeil2008}.
A mechanism leading to gap smearing without invoking any TRI breaking
was proposed 40 years ago in the seminal paper~\cite{LO71}
(see also Ref.~\onlinecite{Oppermann}).
It was shown that the effect of a (phenomenologically introduced)
short-scale disorder in the Cooper attraction constant,
$\lambda = \bar{\lambda} + \delta\lambda(\br)$,
is formally equivalent to the one produced by magnetic impurities \cite{AG}.
Another mechanism \cite{Muttalib87} relates smearing
of the CP with a finite inelastic lifetime of quasiparticles~\cite{Belitz91};
this effect becomes exponentially weak at low
temperatures, $T \ll T_c$. Finally, it has been recently demonstrated
that an apparent DOS smearing seen in tunnelling experiments
may be due to electric fluctuations in the environment~\cite{Pekola10}.

In recent few years an upsurge of interest in experimental studies
of strongly disordered (non-granular) superconductors has been seen,
evidenced, e. g., by Refs.~\cite{Sacepe1,Sacepe2,Astafiev,Aubin,Delft}.
%%%evidenced, e. g., by Refs.~\cite{Sacepe1,Sacepe2,Astafiev,Aubin,Delft,Teun2012}.
Two basic classes of these materials distinguished by the value of the
electron concentration are known (for a review, see Ref.~\cite{Feigelman2010}).
Below we will focus on strongly disordered superconductors with high
(typical metallic) electron density and strong Coulomb
interaction~\cite{Sacepe2,Aubin,Delft},
%%%interaction~\cite{Sacepe2,Aubin,Delft,Teun2012},
where the fermionic mechanism of superconductivity suppression
by disorder~\cite{Finkelstein} is operating.

In this Letter we show that mesoscopic conductance fluctuations~\cite{UCF}
provide a \emph{universal lower bound}\/ for the DOS smearing effects in any
disordered superconducting thin films, effective down to $T=0$.
For the case of thin films
(thickness $d$ is below the low-temperature coherence length $\xi_0$),
the strength of this smearing is completely controlled by the film
dimensionless conductance $g=2\pi\hbar/e^2R_\Box \gg 1$,
and the critical conductance,
$\gcf = \ln^2(\hbar/T_{c0}\tau_*)/2\pi$,
for the fermionic mechanism of superconductivity
suppression~\cite{Finkelstein}.  Here $T_{c0}$ is the transition
temperature in the clean system (we put Boltzmann constant $k_B=1$)
and $\tau_*=\max\{\tau,\tau(d/l)^2\}$, where $\tau$ is the elastic
scattering time and $l=v_F\tau$ is the mean free path.

\begin{figure}
\includegraphics[width=0.8\columnwidth]{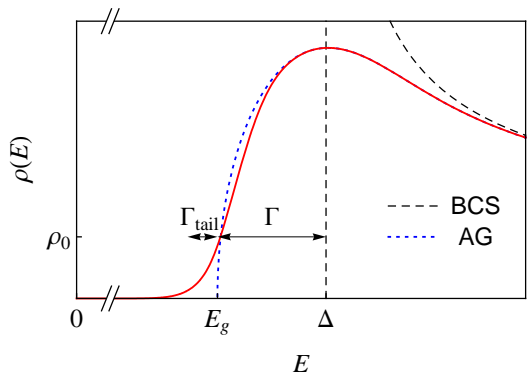}
\caption{(color online). Schematic view of the average DOS
in a dirty superconducting film (solid line).
Broadening of the BCS peak (dashed line) is mainly described
by the semiclassical approximation (dotted line), with
the full DOS containing a significant tail of the subgap states.}
\label{F:DOS}
\end{figure}

We find that the average DOS schematically shown in Fig.~\ref{F:DOS}
is characterized by two energy scales: The width $\Gamma$
measures the broadening of the BCS peak, while $\Gamma_\text{tail}$
determines the exponential decay rate of the subgap DOS,
$\corr{\rho(E)}\propto\exp\{-[(E_g-E)/\Gamma_\text{tail}]^{3/2}\}$.
The tail in the averaged DOS is a manifestation of the local gap
inhomogeneity due to randomness in impurities' configuration.
In the zero-temperature limit, $\Gamma_\text{tail}$
and $\Gamma \gg \Gamma_\text{tail}$ are given by
\be
\label{Gamma-0}
  \frac{\Gamma_\text{tail}}{\Delta_0}
  =
  \left[
    \frac{0.47}{g(g-\gcf)}
  \right]^{2/3} ,
\qquad
  \frac{\Gamma}{\Gamma_\text{tail}}
  =
  \left(\ln\frac{\Delta_0}{\Gamma_\text{tail}}\right)^{2/3}
  .
\ee

The most important feature of the result (\ref{Gamma-0}) is a sharp growth
of the DOS broadening in the vicinity of the quantum critical point, $g=\gcf$.
The same dimensionless parameter, $\delta_d\sim1/g(g-g_c)$, is known
to control the disorder-induced smearing of the thermal transition \cite{SF2005}.

Our quantitative analysis is presented below.

\emph{Mean-field structure of the superconducting state.}---%
We start the analysis of the superconducting state at the mean-field level,
working with disorder-averaged quantities and neglecting spatial fluctuations
of the order parameter.
The effect of the Coulomb interaction on the properties
of disordered superconducting films is usually treated
in terms of an
energy-dependent Cooper amplitude $\lambda(\zeta)$,
with $\zeta=\ln(1/E\tau_*)$ being the logarithm of the running energy scale
(hereafter we set $\hbar=1$).
In the case of the screened Coulomb interaction,
$\lambda(\zeta)$ obeys the RG equation \cite{Finkelstein}:
\be
\label{lambda-RG}
  d\lambda/d\zeta
  =
  \lambda^2
  -
  \lambda_g^2 ,
\qquad
  \lambda_g^2 = 1/2\pi g ,
\ee
where we neglected the triplet sector contribution
and conductance renormalization assuming $\zeta/g\ll1$.
Equation~(\ref{lambda-RG}) describes a competition
between Cooper instability and Coulomb
suppression of Cooper attraction, with the initial condition
$\lambda(0)=\lambda_0$ at $E\sim\tau_*^{-1}$.
The RG flow (\ref{lambda-RG}) drives $\lambda(\zeta)$ to infinity at
$\zeta_*=(2\lambda_g)\ln[(\lambda_0+\lambda_g)/(\lambda_0-\lambda_g)]$,
which signals the superconducting transition with the critical temperature
$T_c=\tau_*^{-1}e^{-\zeta_*}$ \cite{Finkelstein}:
\be
\label{Tc-Fin}
  \frac{T_c\tau_*}{\hbar}
  =
  \left(
    \frac{\sqrt{g}-\sqrt{\gcf}}
         {\sqrt{g}+ \sqrt{\gcf}}
  \right)^{\sqrt{\pi g/2}} .
\ee

Another important feature introduced by the Coulomb interaction
is the energy dependence of the pairing potential $\tilde\Delta(\eps)$
\cite{SF2005}. The latter is defined through the quasiclassical
Gorkov function in the Matsubara representation,
$F(\eps)=\tilde\Delta(\eps)/[\eps^2+\tilde\Delta^2(\eps)]^{1/2}$.
The function $\tilde\Delta(\eps)$
can be obtained from the self-consistency equation (SCE)
\be
\label{SCE}
  \tilde\Delta(\eps)
  =
  \pi T \sum_{\eps'} \lambda(\eps,\eps') F(\eps') ,
\ee
where $\eps$ is the fermionic Matsubara energy,
and the energy-dependent Cooper amplitude is given by
\be
\label{lambda(E)}
  \lambda(\eps,\eps')
  =
  \lambda_0
  -
  \lambda_g^2 \ln[1/\max(\eps,\eps')\tau_*]
  .
\ee

A description based on Eqs.~(\ref{SCE}) and (\ref{lambda(E)}),
where all energies are retained, is alternative to successive
elimination of high-energy degrees of freedom by the RG evolution of $\lambda(\zeta)$.
The logarithmic correction to $\lambda_0$ in Eq.~(\ref{lambda(E)})
corresponds to the last term in RG equation (\ref{lambda-RG}).
At the same time, the first term in Eq.~(\ref{lambda-RG})
is automatically taken into account by the summation
over energy in Eq.~(\ref{SCE}).
An approach based on Eqs.~(\ref{SCE}) and (\ref{lambda(E)})
is equivalent to the RG in the determination of $T_c$, but can be used also
to find $\tilde\Delta(\eps)$ in the superconducting phase at $T<T_c$.

Equation (\ref{SCE}) generalizes the SCE in the BCS theory.
To find the energy dependence of $\tilde\Delta(\eps)$ at large energies,
$\eps\gtrsim T_c$, we rewrite Eq.~(\ref{SCE}) as a linear integral
equation in terms of the logarithmic variable $\zeta$,
which is readily solved by reducing to a differential equation
owing to a simple form of the kernel (\ref{lambda(E)}).
As a result, we arrive at
\be
\label{Delta(E)}
  \tilde\Delta(\eps)
  =
  \Delta(T)
  \bigl[
    (\eps/T_c)^{\lambda_g}
    +
    (\eps/T_c)^{-\lambda_g}
  \bigr]/2
  ,
\ee
valid for $\eps\gtrsim T_c$.
Here $\Delta(T)$ is a function of temperature which should be determined
from the full equation (\ref{SCE}), where the region of small energies,
$\eps\sim T_c$, becomes important. Neglecting a slow
$\eps$-dependence of $\tilde\Delta(\eps)$, we conclude that $\Delta(T)$
is related to $T_c$ as in the standard BCS theory [in particular,
$\Delta(0)=1.76 T_c$].

According to Eq.~(\ref{Delta(E)}), high-energy ($E\gg T_c$)
electrons experience a larger value of the effective pairing
potential \cite{SF2005}. This effect is most pronounced
in the limit of strong suppression of superconductivity, $T_c\ll T_{c0}$,
when the overall enhancement becomes large: $\tilde\Delta(\tau_*^{-1})\gg\Delta(T)$.

\emph{Smearing by inhomogeneities.}---%
The mean-field theory developed above describes disorder-averaged
quantities. In the presence of a quenched disorder, the order
parameter becomes non-uniform and the sharp BCS peak gets broadened.
Analytical description of this effect is complicated
due to the failure of the perturbation theory at $E\to\Delta$.
Earlier experience \cite{LO71,MeyerSimons2001} suggests that
the problem can be conveniently tackled in two steps:
\begin{enumerate}
\item
First one has to find the correlation function
$\corr{\Delta(\br)\Delta(\br')}$.
Since this is a thermodynamic quantity involving contributions
from many energies, it can be obtained in the Matsubara
representation by a regular perturbation theory.
\item
Then behavior of electrons in the field of a spatially
fluctuating $\Delta(\br)$ can be considered independently
for each real energy $E$.
\end{enumerate}
This scheme based on the perturbation
theory is applicable provided that disorder smearing is small.

In dirty superconductors, diffusive motion of electrons
is described by the Usadel equation \cite{Usadel}
supplemented by the SCE (\ref{SCE}). Various types of disorder,
such as magnetic impurities \cite{AG,LamacraftSimons}
or fluctuating coupling constant \cite{LO71,MeyerSimons2001},
can be easily incorporated into the scheme as random fields
in the Usadel equation.
The situation with \emph{universal mesoscopic disorder}\
(intrinsic fluctuations of the potential disorder)
we are considering is different:
Since the Usadel equation is already written
for the ensemble-averaged quantities, mesoscopic potential
disorder cannot be included there as some extra fluctuating field.
To find the correlation function $\corr{\Delta(\br)\Delta(\br')}$,
one has to go beyond the Usadel equation, consider
\emph{two replicas}\ of the system
and average over soft diffusive modes \cite{SF2005},
similar to calculation of
the universal conductance fluctuations
(UCF) \cite{UCF}.

\emph{Mesoscopic fluctuations of the order parameter.}---%
In the presence of disorder, the SCE (\ref{SCE})
contains two sources of disorder: mesoscopic fluctuations
of the coupling constant, $\lambda_\text{dis}(\eps,\eps';\br)$
(in its Coulomb part), and mesoscopic fluctuations
of the Gorkov function, $F_\text{dis}(\eps;\br)$.
These quantities exhibit fluctuations even for a uniform
order parameter since they are governed by diffusive
motion of electrons sensitive to mesoscopic disorder.
Due to the SCE, this fluctuations will
result in an inhomogeneous contribution to the order parameter:
$\tilde\Delta(\eps;\br) = \tilde\Delta(\eps) + \tilde\Delta_1(\eps;\br)$.
The latter, in turn, will modify $F$ which therefore can be
represented in the form
$F = F_0 + (\partial F_0/\partial \tilde\Delta)\tilde\Delta_1 + F_\text{dis}$.
Substituting this in the SCE (\ref{SCE})
and linearizing we get an equation for $\tilde\Delta_1(\eps)$
in the Fourier representation:
\begin{multline}
\label{Delta1-eq}
  \tilde\Delta_1(\eps,\bq)
  -
  \pi T \sum_{\eps'}
  \lambda(\eps,\eps')
  \frac{\partial F_0(\eps',\bq)}{\partial\tilde\Delta(\eps',\bq)}
  \tilde\Delta_1(\eps',\bq)
\\{}
  =
  \pi T \sum_{\eps'}
  \left[
    \lambda(\eps,\eps') F_\text{dis}(\eps',\bq)
    +
    \lambda_\text{dis}(\eps,\eps',\bq) F_0(\eps')
  \right]
  .
\end{multline}
In the 2D case, the main contribution to the CP smearing
comes from large scales [see Eq.~(\ref{eta})], much exceeding
the correlation length $r_c\sim\xi_0\sim\sqrt{D/T_c}$
for mesoscopic fluctuations of $F_\text{dis}$ and $\lambda_\text{dis}$.
Therefore it suffices to consider $\tilde\Delta_1(\eps)$
at zero momentum which will be implied below.

Inverting the matrix in the left-hand side of Eq.~(\ref{Delta1-eq}),
we express $\tilde\Delta_1(\eps)$
in terms of $F_\text{dis}$ and $\lambda_\text{dis}$.
To study the CP smearing we need the small-energy limit
of $\tilde\Delta_1(\eps)$ with $\eps\sim T_c$, where the effect
of fluctuations is to modify $\Delta(T) \to \Delta(T)+\delta\Delta(T)$,
with $\delta\Delta(T)$ given by \cite{Supplementary}
\begin{multline}
\label{Delta1-res}
  \delta\Delta(T)
  =
  L_0 \Bigl(\frac{T}{T_c}\Bigr)
  \frac{(2\pi T)^2}{\Delta(T)}
  \sum_{\eps_1,\eps_2>0}
  F_0(\eps_1)
\\{}
  \times
  \left[
    \lambda(\eps_1,\eps_2) F_\text{dis}(\eps_2)
    +
    \lambda_\text{dis}(\eps_1,\eps_2) F_0(\eps_2)
  \right]
  .
\end{multline}
Here $L_0$ is the fluctuation propagator at zero momentum and frequency
in the BCS theory:
\be
\label{L0}
  L_0^{-1}\Bigl(\frac{T}{T_c}\Bigr)
  =
  \pi T \sum_\eps
  \frac{\Delta^2(T)}{\mathfrak{E}^3}
  ,
\qquad
\mathfrak{E}=\sqrt{\eps^2+\Delta^2(T)} ,
\ee
with the asymptotic behavior $L_0\approx 4\pi^2T^2/7\zeta(3)\Delta^2(T)$ at $T\to T_c$,
and $L_0=1$ at $T=0$.

\begin{figure}
\includegraphics[width=0.8\columnwidth]{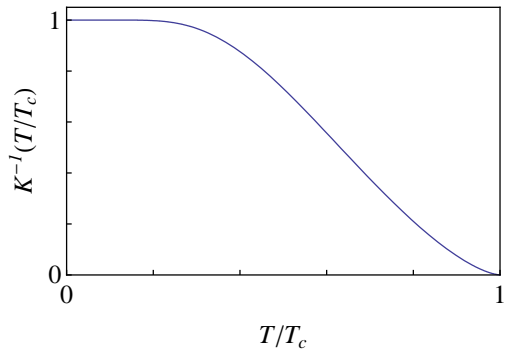}
\caption{(color online). Plot of the function $K^{-1}(T/T_c)$,
where $K(t)=L_0^2(t) N(t)$ is defined by Eqs.~(\ref{L0}) and (\ref{N0}).}
\label{F:K}
\end{figure}

Therefore, smearing of the CP and behavior near
the gap edge are determined by a single number,
$f(0) = \corr{\delta\Delta\delta\Delta}_{\bq=0}$,
which can be easily obtained from the correlation functions
of $F_\text{dis}$ and $\lambda_\text{dis}$ with the help
of Eq.~(\ref{Delta1-res}).
Since the Coulomb correction to $\lambda$ already contains
a closed loop (return probability), the correlation functions
$\corr{F_\text{dis}F_\text{dis}}$,
$\corr{F_\text{dis}\lambda_\text{dis}}$ and
$\corr{\lambda_\text{dis}\lambda_\text{dis}}$
are given by the \mbox{one-,} two- and three-loop diagrams
in soft diffusive modes, respectively.
The overall contribution is given by \cite{Supplementary}
\be
\label{f(0)}
  f(0)
  =
  \frac{\pi D\Delta(T)}{g(g-\gcf)}
  K\Bigl(\frac{T}{T_c}\Bigr)
  ,
\ee
where $K(T/T_c)=L_0^2(T/T_c)N(T/T_c)$ (see Fig.~\ref{F:K}), and
\be
\label{N0}
  N\Bigl(\frac{T}{T_c}\Bigr)
  =
  16 T^2 \sum_{\eps_1,\eps_2>0}
  \frac{\Delta(T)}{\mathfrak{E}_1 \mathfrak{E}_2 (\mathfrak{E}_1+\mathfrak{E}_2)}
  ,
\ee
with $N\approx 14\zeta(3)\Delta(T)/\pi^3 T$ at $T\to T_c$,
and $N=1$ at $T=0$.
Replacement of $1/g^2$ by $1/g(g-\gcf)$ in Eq.~(\ref{f(0)}) is due to
high-energy contributions, $T_c<\eps<\tau_*^{-1}$, where $\tilde\Delta(\eps)$
is enhanced according to Eq.~(\ref{Delta(E)}). In the limit of strong
$T_c$ suppression, $g-\gcf<\gcf$, the leading source of disorder comes
from mesoscopic fluctuations of the return probability
in $\corr{\lambda_\text{dis}\lambda_\text{dis}}$.
Equation (\ref{f(0)}) is consistent with our previous result
in the vicinity of $T_c$ \cite{SF2005},
generalizing it to arbitrary temperatures $T<T_c$.

\emph{Mean-field density of states.}---%
The average DOS, $\corr{\rho(E)}=\rho_0\Re\corr{\cos\theta(E,\br)}$,
is expressed in terms of the spectral angle $\theta$
which satisfies the Usadel equation,
$(D/2) \nabla^2 \theta + iE \sin\theta + \Delta(\br) \cos\theta = 0$,
with a random order parameter, $\Delta(\br)=\Delta_0+\delta\Delta(\br)$
[here $\Delta_0\equiv\Delta(T)$].
Integrating out short-range degrees of freedom one gets an
equation for the long-range behavior of $\theta(E,\br)$
\cite{LO71,MeyerSimons2001}:
\be
\label{Usadel-eta}
  \frac{D}{2} \nabla^2 \theta
+ iE \sin\theta
+ \Delta(\br) \cos\theta
  -
  \Delta_0 \eta \cos\theta \sin\theta
  =
  0
  ,
\ee
where the depairing strength is expressed in terms of the correlation function
$f(\bq) = \corr{\delta\Delta \delta\Delta}_{\bq}$ as
\be
\label{eta}
  \eta
  =
  \frac{2}{\Delta_0}
  \int
  \frac{f(\bq)}{Dq^2}
  \frac{d^2\bq}{(2\pi)^2}
  .
\ee
This expression has been originally derived
in Ref.~\cite{LO71} for the 3D geometry.
The last term in Eq.~(\ref{Usadel-eta})
coincides with the depairing term due to magnetic impurities
derived by Abrikosov and Gorkov (AG) \cite{AG}.
It leads to the broadening of the CP (shown by the dotted
line in Fig.~\ref{F:DOS}), with the hard gap at
$E_g^\text{AG} = (1-\eta^{2/3})^{3/2} \Delta_0$.

In the marginal 2D case, mapping to the problem of magnetic impurities
should be done with care. Contrary to the 3D geometry, now the integral
in Eq.~(\ref{eta}) is logarithmically divergent at small $q$.
An appropriate
cutoff can be established by retaining the Cooperon mass in Eq.~(\ref{eta}):
$Dq^2\mapsto Dq^2 + D/L_E^2$, with $D/L_E^2 =
2 ( - iE \cos\theta + \Delta_0 \sin\theta + \Delta_0 \eta \cos2\theta)$.
Thus, in the 2D geometry, $\eta$ becomes a function of $\theta$ and $E$,
making the depairing term in the Usadel equation more complicated
than the simple AG term. However since the dependence of $\eta$ on $L_E$
is  logarithmically slow, we can evaluate it at the AG solution
replacing $\theta$ by $\theta_\text{AG}(E)$:
\be
\label{L(E)}
  L_E
  =
  \frac{D^{1/2}}{[24\Delta_0(E_g-E)]^{1/4}}
  \sim
  \xi(T) \left( \frac{E_g}{E_g-E} \right)^{1/4}
\ee
(we assume $\eta\ll1$).
As a result, the depairing factor becomes energy-dependent:
\be
\label{eta-2D}
  \eta(E)
  =
  \frac{K(T/T_c)}{g(g-\gcf)}
  \ln\frac{\min(L_E,L_g)}{\xi_0}
  ,
\ee
where we had to introduce an infrared length scale $L_g$ to
regularize the otherwise divergent $\eta(E\to E_g)$.
Its appearance is related to the breakdown
of the mean-field approximation in the narrow region
$|E-E_g|\lesssim \Gamma_\text{tail}$ \cite{LO71}, where proliferation
of instantons generates a finite correlation length
$L_g \sim \xi(T) (E_g/\Gamma_\text{tail})^{1/4}$.
Substituting $\eta=\eta(E_g)$ into $E_g^\text{AG}$,
we obtain for $\Gamma \equiv \Delta(T)-E_g$:
\be
\label{Gamma}
  \frac{\Gamma}{\Delta(T)}
  =
  \frac{3}{2}
  \left[
    \frac{K(T/T_c)}{4g(g-\gcf)}
    \ln\frac{L_g(T)}{\xi_0}
  \right]^{2/3}
  ,
\ee
which in the zero-temperature limit reduces to Eq.~(\ref{Gamma-0}).

The theory developed above applies to quasi-2D films
with the thickness $d\ll\xi_0$. For finite $d\lesssim\xi_0$,
there exists a contribution to $\eta$ coming from short
scales ($l\ll r\ll d$) where electron diffusion
is 3D \cite{Supplementary}:
\be
\label{eta3D-res}
  \eta_\text{3D}
  =
  \frac{2\Delta(T)}{\pi^2 \hbar D}
  \left( \frac{\rho}{R_Q} \right)^2
  \ln\frac dl
  \sim
  \frac{d^2}{\xi^2(T)}
  \frac{1}{g^2}
  \ln\frac dl
  ,
\ee
where $\rho$ is the film resistivity, $R_Q=2\pi\hbar/e^2$,
and $\xi(T)$ is the temperature-dependent coherence length.
Correction (\ref{eta3D-res}) which should be added to Eq.~(\ref{eta-2D})
leads to a small increase of the width $\Gamma$.

\emph{Subgap states.}---%
A hard gap in the excitation spectrum predicted by the AG theory
is smeared by disorder leading to the formation of states at $E<E_g$.
These localized states are identified as instantons in the nonlinear
equation (\ref{Usadel-eta}) \cite{LO71,LamacraftSimons,MeyerSimons2001,OSF01}.
In Ref.~\cite{LO71}, Lifshitz-type arguments were used to determine
an optimal fluctuation of $\Delta(\br)$ in Eq.~(\ref{Usadel-eta})
giving rise to a finite DOS at $E<E_g$.
On the contrary, in Ref.~\cite{MeyerSimons2001}, Eq.~(\ref{Usadel-eta})
was considered for a uniform $\Delta(\br)=\Delta_0$, and instantons
related with intrinsic nonlinearity of the Usadel equation were analyzed.
The results of Refs.~\cite{LO71} and \cite{MeyerSimons2001} are different and
represent two asymptotics of a unique function of $E_g-E$
(a detailed discussion will be given elsewhere~\cite{SF-inprep}).
For small $E_g-E$, nonlinearity is weak and the subgap DOS is due to
optimal fluctuations of $\Delta(\br)$ \cite{LO71}, whereas the mechanism
of Ref.~\cite{MeyerSimons2001} is applicable only for very large $E_g-E$,
where the DOS is exponentially small.

Generalizing the 3D analysis of Ref.~\cite{LO71}
to the 2D case \cite{Supplementary},
we find that the DOS decays exponentially in the subgap region:
\begin{gather}
\label{DOS-tail}
  \corr{\rho(E)}
  \propto
  \exp\bigl\{
    -
    [(E_g-E)/\Gamma_\text{tail}]^{3/2}
  \bigr\} ,
\\
\label{Gamma-tail}
  \frac{\Gamma_\text{tail}}{\Delta(T)}
  =
  \left[
    0.47
    \frac{K(T/T_c)}{g(g-\gcf)}
  \right]^{2/3}
\end{gather}
[in the $d$-dimensional space, $\ln\corr{\rho(E)}\propto-(E_g-E)^{2-d/4}$].
At $T=0$, Eq.~(\ref{Gamma-tail}) reduces to Eq.~(\ref{Gamma-0}).
Equation (\ref{DOS-tail}) is valid as long as $E_g-E>\Gamma_\text{tail}$.
Note that the instanton action obtained in Ref.~\cite{MeyerSimons2001}
neglecting $\Delta(\br)$ fluctuations is extremely large, $S\sim g$,
already at $E_g-E\sim\Gamma_\text{tail}$.
Hence, $\corr{\rho(E)}$ follows Eq.~(\ref{DOS-tail})
for all conceivable $E<E_g$.

\emph{Discussion.}---%
Superconducting samples always have some amount of disorder
which leads to smearing of the BCS density of states.
We have considered the case of a minimal possible disorder ---
intrinsic randomness in a homogeneously disordered film
responsible for the UCF in the normal state.
The average DOS sketched in Fig.~\ref{F:DOS} is characterized
by two energy scales.
The shift of the gap edge $\Delta_0 \to E_g$
and related suppression of the CP height is controlled by
the parameter $\Gamma$,
so that $\rho_\text{max}(E)/\rho_0 \approx (\Delta/\Gamma)^{1/2}$.
At the same time, the width of the subgap tail is determined
by a different parameter $\Gamma_\text{tail} < \Gamma$.
Both $\Gamma$ and $\Gamma_\text{tail}$ are small in a clean system
but get enhanced as the film becomes less conductive approaching
the critical point, $g\to\gcf$. Smearing of the DOS structure
becomes very strong at $g -\gcf \sim 1/\gcf$, where our theory
becomes inapplicable.
The CP broadening is temperature dependent and grows at $T\to T_c$
due to the growth of the function $K(T/T_c)$ (see Fig.~\ref{F:K}).

Our results for the disorder-induced DOS smearing should be compared
with the smearing due to inelastic scattering~\cite{Muttalib87}.
In a 2D system, the inelastic rate due to Coulomb interaction
is of the order of $\Gamma_{ee}\sim T/g$ at $T\sim T_c$,
and gets exponentially suppressed,
$\Gamma_{ee}\propto e^{-\Delta(T)/T}$, at $T\ll T_c$ \cite{Belitz91}.
Therefore our mechanism always dominates at low temperatures.
It also always dominates close to $T_c$, where $K(T/T_c) \propto (1-T/T_c)^{-3/2}$.
The general relation between the rates $\Gamma$ and $\Gamma_{ee}$
depends on the proximity to the quantum critical point.
Relatively far from it, at $g-\gcf\gtrsim\sqrt\gcf$,
inelastic scattering is the leading source of the DOS smearing
at intermediate temperatures, $T_1 \lesssim T \lesssim T_2$,
where $T_1\sim\Delta_0/\ln\beta$ and $T_c-T_2\sim\beta^{-2}$,
with $\beta=(g-\gcf)^{2/3}/g^{1/3}$.
For films closer to criticality, $g-\gcf\lesssim\sqrt\gcf$,
disorder-induced smearing always dominates over the inelastic smearing.

Developed theory is expected to be most appropriate for very thin films of
amorphous metallic superconductors (e.g., Mo-Ge, Nb-Si, W-Re, Nb$_3$Ge),
where suppression of $T_c$ by disorder is described by
the fermionic mechanism~\cite{Finkelstein}, and $R_\Box \geq 1$--2 k$\Omega$.

Finally, we emphasize that the widely used phenomenological Dynes ansatz,
$\rho(E) = \rho_0\Re (E-i\Gamma)/[(E-i\Gamma)^2-\Delta^2]^{1/2}$
\cite{Dynes} is inapplicable when disorder is the
main source of the broadening. The actual DOS profile then depends
on two energy parameters, $\Gamma$ and $\Gamma_\text{tail}$,
and decays exponentially rather than algebraically at $E<E_g$.

We are grateful to I. S. Burmistrov, T. M. Klapwijk, and A. Silva
for useful discussions.
This work was supported by the RFBR grant No.\ 10-02-01180,
the Dynasty Foundation, and the Russian Federal Agency
of Education (contract No.\ P799) (M. S.).

%%%%%%%%%%%%%%%%%%%%%%%%%%%%%%%%%%%%%%%%%%%%%%%%%%%%%

\onecolumngrid

\clearpage

\renewcommand{\theequation}{S\arabic{equation}}

\def\br{{\bf r}}
\def\bp{{\bf p}}
\def\bq{{\bf q}}
\def\bk{{\bf k}}
\def\bv{{\bf v}}

\def\arccosh{\mathop{\rm arccosh}}
\def\eps{\varepsilon}
\def\gcf{g_{c\text{F}}}

\newcommand{\const}{\text{const}}

\centerline{\large\textbf{Supplemental Material}}

\vskip 12mm

\section{I. Inversion of the fluctuation propagator}

Here we invert the matrix fluctuation propagator defined by the left-hand
side of Eq.~(\ref{Delta1-eq}):
\be
\label{Lambda-f}
  \tilde\Delta_1(\eps)
  -
  \pi T \sum_{\eps'}
  \lambda(\eps,\eps')
  \frac{\partial F_0(\eps')}{\partial\tilde\Delta(\eps')}
  \tilde\Delta_1(\eps')
  =
  \phi(\eps) .
\ee
Evaluating the derivative of
$F(\eps')=\tilde\Delta(\eps')/[\eps'^2+\tilde\Delta^2(\eps')]^{1/2}$, we get
\be
\label{Lambda-f2}
  \tilde\Delta_1(\eps)
  -
  \pi T \sum_{\eps'}
  \lambda(\eps,\eps') F_0(\eps') \frac{\tilde\Delta_1(\eps')}{\tilde\Delta(\eps')}
  +
  \pi T \sum_{\eps'}
  \frac{\lambda(\eps,\eps')\tilde\Delta^2(\eps')}{[\eps'^2+\tilde\Delta^2(\eps')]^{3/2}}
  \tilde\Delta_1(\eps')
  =
  \phi(\eps) .
\ee
Since both $\tilde\Delta(\eps')$ and $\tilde\Delta_1(\eps')$
are logarithmically slow functions of $\eps'$, the second sum
can be easily evaluated and we arrive at
\be
\label{Lambda-f3}
  \tilde\Delta_1(\eps)
  -
  \pi T \sum_{\eps'}
  \lambda(\eps,\eps') F_0(\eps') \frac{\tilde\Delta_1(\eps')}{\tilde\Delta(\eps')}
  +
  L_0^{-1}
  \lambda(\eps,T_c)
  \tilde\Delta_1(T_c)
  =
  \phi(\eps) ,
\ee
where $L_0$ is the fluctuation propagator at zero frequency and momentum
in the BCS theory:
\be
\label{L0-supp}
  L_0^{-1}\left(\frac{T}{T_c}\right)
  =
  \pi T \sum_{\eps}
  \frac{\Delta^2(T)}{[\eps^2+\Delta^2(T)]^{3/2}}
  =
  \begin{cases}
  \displaystyle
  \frac{7\zeta(3)\Delta^2(T)}{4\pi^2 T^2}
  ,
  & T_c-T\ll T_c ;
  \\[6pt]
  \displaystyle
  1
  ,
  & T \ll T_c .
  \end{cases}
\ee

The value of $\tilde\Delta_1(T_c)$ can be easily obtained
from Eq.~(\ref{Lambda-f3}). In order to do this we multiply it
by $F(\eps)$ and sum over $\eps$. Using the SCE (\ref{SCE}),
we immediately see that the first two terms in Eq.~(\ref{Lambda-f3})
cancel and we obtain
\be
\label{Delta1*}
  \tilde\Delta_1(T_c)
  =
  \frac{L_0}{\Delta(T)} \,
  \pi T \sum_{\eps} F_0(\eps)
  \phi(\eps) ,
\ee
where we use that, according to Eq.~(\ref{Delta(E)}),
$\tilde\Delta(T_c) = \Delta(T)$.

\section{II. Correlation function $\corr{\Delta_1\Delta_1}$ due to mesoscopic fluctuations}

In this Section we evaluate the zero-momentum correlation function
\be
\label{Phi-def}
  \Phi
  =
  \frac{\corr{\tilde\Delta_1(T_c)\tilde\Delta_1(T_c)}_{\bq=0}}
    {\tilde\Delta(T_c)\tilde\Delta(T_c)}
\ee
due to mesoscopic fluctuations of $F_\text{dis}$ and $\lambda_\text{dis}$.

\subsection{Correlation function $\corr{F_\text{dis}F_\text{dis}}$}

The correlator of Gorkov functions in the Matsubara representation
is calculated with the help of imaginary-time replica sigma-model
following the line of Ref.~[S1]. The resulting expression
has the form
\be
  \label{<FF>}
  \corr{
  F_\text{dis}(\eps,\br)
  F_\text{dis}(\eps',\br')
  }
  =
  F_0(\eps)
  F_0(\eps')
  \frac{[\Pi_{\eps\eps'}(\br,\br')]^2}{(\pi\nu)^2}
  ,
\ee
where $\nu$ is the 2D one-particle DOS at the Fermi level (per single
spin projection), and $\Pi$ is the diffusion operator on top
of the superconducting state:
\be
\label{Pi-1}
  \Pi^{-1}_{\eps\eps'}
  =
  - D \nabla^2 + \mathfrak{E}(\eps) + \mathfrak{E}(\eps') ,
\ee
where
\be
\label{sqrt}
  \mathfrak{E}(\eps)
  =
  \sqrt{\eps^2+\tilde\Delta^2(\eps)} .
\ee
For the zero Fourier component we get
\be
  \corr{F_\text{dis}(\eps) F_\text{dis}(\eps')}_{\bq=0}
  =
  \frac{F_0(\eps)F_0(\eps')}{4\pi^3\nu^2D}
  \frac{1}{\mathfrak{E}(\eps)+\mathfrak{E}(\eps')} .
\ee

The corresponding contribution to Eq.~(\ref{Phi-def}) has the form
\be
\label{PhiFF1}
  \Phi^{(FF)}
  =
  \frac{L_0^2}{\Delta^4(T)}
  \,
  \frac{(2\pi T)^4}{4\pi^3\nu^2D}
  \sum_{\eps_1,\eps_2,\eps_3,\eps_4>0}
  \lambda(\eps_1,\eps_2)
  \lambda(\eps_3,\eps_4)
  \frac{F_0(\eps_1) F_0(\eps_2) F_0(\eps_3) F_0(\eps_4)}
    {\mathfrak{E}(\eps_2)+\mathfrak{E}(\eps_4)}.
\ee

This expression is typical to fluctuation contributions to $\Phi$.
Among four energy summations, two are logarithmic involving large
energies, $\eps\gg T_c$, while the other two come from
$\eps\sim T_c$, where $\tilde\Delta(\eps)$ can be approximated by $\Delta(T)$.
The latter summations introduce the dimensionless function
\be
\label{N}
  N\left(\frac{T}{T_c}\right)
  =
  16 T^2 \sum_{\eps_1,\eps_2>0}
  \frac{\Delta(T)}{\mathfrak{E}_1 \mathfrak{E}_2 (\mathfrak{E}_1+\mathfrak{E}_2)}
  =
  \frac{2}{\pi}
  \int_0^\infty
  \frac{d\theta}{\cosh\theta}
  \tanh^2\left[\frac{\Delta(T)}{2T}\cosh\theta\right]
  =
  \begin{cases}
  \displaystyle
  \frac{14\zeta(3)}{2\pi^3} \frac{\Delta(T)}{T} ,
  & T_c-T \ll T_c ;
  \\[10pt]
  \displaystyle
  1 ,
  & T \ll T_c .
  \end{cases}
\ee

Performing summations over $\eps_2$ and $\eps_4$ in Eq.~(\ref{PhiFF1}) we get
\be
\label{PhiFF2}
  \Phi^{(FF)}
  =
  \frac{1}{16\pi\nu^2D\Delta^3(T)}
  L_0^2\left(\frac{T}{T_c}\right)
  N\left(\frac{T}{T_c}\right)
  \left(
    2\pi T \sum_{\eps>0} \lambda(\eps,T_c) F_0(\eps)
  \right)^2
\ee
Summation is done with the help of the SCE (\ref{SCE}),
and we obtain finally
\be
\label{Phi(FF)}
  \Phi^{(FF)}
  =
  \frac{1}{16\pi\nu^2D\Delta(T)}
  L_0^2\left(\frac{T}{T_c}\right)
  N\left(\frac{T}{T_c}\right)
  .
\ee

\subsection{Correlation function $\corr{\delta\lambda_\text{dis}\delta\lambda_\text{dis}}$}

Mesoscopic fluctuations of the return probability which determines
$\delta\lambda(\eps,\eps') = -\lambda_g^2 \ln[1/\max(\eps,\eps')\tau]$
have been calculated for the normal state in Ref.~[S2]:
\be
\label{ll-corr}
  \corr{\lambda_\text{dis}(\eps_1,\eps_2)\lambda_\text{dis}(\eps_3,\eps_4)}_{\bq=0}
  =
  \frac{\delta\lambda(\eps_1,\eps_2)\delta\lambda(\eps_3,\eps_4)}
    {16\pi^3\nu^2D}
  \left(
    \frac{1}{|\eps_{13}|}
    +
    \frac{1}{|\eps_{14}|}
    +
    \frac{1}{|\eps_{23}|}
    +
    \frac{1}{|\eps_{24}|}
  \right) ,
\ee
where $\eps_{ij}\equiv\eps_i+\eps_j$.

Generalization to the superconducting case is achieved by replacing
$\eps \to \mathfrak{E}(\eps)$. All four terms in Eq.~(\ref{ll-corr})
equally contribute to $\Phi$, and we get
\be
  \Phi^{(\lambda\lambda)}
  =
  \frac{L_0^2}{\Delta^4(T)}
  \,
  \frac{(2\pi T)^4}{4\pi^3\nu^2D}
  \sum_{\eps_1,\eps_2,\eps_3,\eps_4>0}
  \delta\lambda(\eps_1,\eps_2)
  \delta\lambda(\eps_3,\eps_4)
  \frac{F_0(\eps_1) F_0(\eps_2) F_0(\eps_3) F_0(\eps_4)}
    {\mathfrak{E}(\eps_2)+\mathfrak{E}(\eps_4)}
  .
\ee
Analogously to Eq.~(\ref{PhiFF1})], summations over $\eps_2$ and
$\eps_4$ yields $N(T/T_c)$, while summations over $\eps_1$ and $\eps_3$
are converted to a logarithmic integral:
\be
  \Phi^{(\lambda\lambda)}
  =
  \frac{1}{16\pi\nu^2D\Delta(T)}
  L_0^2\left(\frac{T}{T_c}\right)
  N\left(\frac{T}{T_c}\right)
  \left(
  \int_0^{\zeta_*}
  \delta\lambda(\zeta,\zeta_*) \cosh\lambda_g(\zeta_*-\zeta) \, d\zeta
  \right)^2
  ,
\ee
and hence
\be
\label{Phi(ll)}
  \Phi^{(\lambda\lambda)}
  =
  \frac{1}{16\pi\nu^2D\Delta(T)}
  L_0^2\left(\frac{T}{T_c}\right)
  N\left(\frac{T}{T_c}\right)
  [\cosh\lambda_g\zeta_*-1]^2
  .
\ee

\subsection{Correlation function $\corr{F_\text{dis}\delta\lambda_\text{dis}}$}

The cross term is evaluated analogously:
\be
\label{Phi(Fl)}
  \Phi^{(F\lambda)}
  =
  -2
  \frac{1}{16\pi\nu^2D\Delta(T)}
  L_0^2\left(\frac{T}{T_c}\right)
  N\left(\frac{T}{T_c}\right)
  [\cosh\lambda_g\zeta_*-1]
  .
\ee

\subsection{Resulting expression for $\corr{\Delta_1\Delta_1}$}

Adding (\ref{Phi(FF)}), (\ref{Phi(ll)}) and (\ref{Phi(Fl)}), we obtain
\be
\label{Phi-res}
  \Phi
  =
  \frac{1}{16\pi\nu^2D\Delta(T)}
  K\Bigl(\frac{T}{T_c}\Bigr)
  \cosh^2\lambda_g\zeta_*
  =
  \frac{\pi D}{g(g-g_c)\Delta(T)}
  K\Bigl(\frac{T}{T_c}\Bigr)
  ,
\ee
where we have introduced $K(T/T_c)=L_0^2(T/T_c)N(T/T_c)$,
used the relation $g=4\pi\nu D$, and employed
$\cosh^2\lambda_g\zeta_*=g/(g-g_c)$.
With the help of Eq.~(\ref{Phi-def}), one readily obtains
Eq.~(\ref{f(0)}) of the main text.

It's worth noting that Eq.~(\ref{Phi-res}) agrees with our previous
results [S2]. Indeed, in the limit $T\to T_c$, Eq.~(\ref{Phi-res})
can be simplified with the help of Eqs.~(\ref{L0-supp}) and (\ref{N}) as
\be
\label{Delta-corr}
  \ccorr{\Delta(T)\Delta(T)}_{\bq=0}
  =
  \frac{2T^3}{7\zeta(3)\nu^2D\Delta^2(T)}
  \cosh^2\lambda_g\zeta_* .
\ee
On the other hand, the same correlator can be obtained from
the correlation function of the coefficient $\alpha$
in the Ginzburg-Landau (GL) expansion [S2]:
\be
  \ccorr{\alpha\alpha}_{\bq=0}
  =
  \frac{7\zeta(3)}{8\pi^4DT} \cosh^2\lambda_g\zeta_*
\ee
with the help of the relation
\be
\label{DD-old}
  \ccorr{\Delta(T)\Delta(T)}_{\bq=0}
  =
  \frac{\ccorr{\alpha\alpha}_{\bq=0}}{4\beta^2\Delta^2(T)} ,
\ee
where $\beta=7\zeta(3)\nu/(8\pi^2T^2)$ is the nonlinear
coefficient in the GL functional.
One can easily verify that Eq.~(\ref{DD-old}) coincides
with Eq.~(\ref{Delta-corr}).

\section{III. Density of the subgap states}

Introducing $\theta=\pi/2+i\psi$, we rewrite Eq.~(\ref{Usadel-eta}) as
\be
\label{Usadel-psi}
- \xi^2 \nabla^2 \psi + F(\psi)
  =
- \frac{\delta\Delta(\br) \sinh\psi}{\Delta_0}
  ,
\ee
where $\xi^2=D/2\Delta_0$, and
\be
  F(\psi)
  =
- \frac{E}{\Delta_0} \cosh\psi
+ \sinh\psi
- \eta \sinh\psi \cosh\psi .
\ee

At the minigap ($E=E_g$), $\cosh\psi_g=\eta^{-1/3}$.
For small deviation from the gap, the function $F(\psi)$
in the vicinity of its maximum can be written as
\be
\label{Omega-rho}
  F(\psi) \approx \Omega (\psi-\psi_0) - \rho (\psi-\psi_0)^2
  ,
\ee
where
\be
  \Omega = \sqrt{6}
%  (1-\eta^{2/3})^{1/4}
  \sqrt[4]{1-\eta^{2/3}}
  \sqrt{\frac{E_g-E}{\Delta_0}}
  ,
\qquad
  \rho = \frac32 \eta^{1/3} \sqrt{1-\eta^{2/3}}
  .
\ee
Comparison of the linear in $\psi$ terms in the left-hand side
of Eq.~(\ref{Usadel-psi}) defines the relevant length scale
\be
\label{SM:L(E)}
  L_E = \frac{\xi}{\sqrt\Omega} .
\ee

At the mean-field level, $\psi$ is real below the gap.
Finite DOS corresponds to appearance of a nonzero $\Im\psi$
due to a large negative fluctuation of $\delta\Delta(\br)$.
Its probability is given by
\be
  {\cal P}
  \propto
  \exp\left(
  - \frac{1}{2f(0)} \int \delta\Delta^2(\br) \, d^dr
  \right)
  \equiv
  e^{-S_d} ,
\ee
where we have used that the instanton size, $L_E$, exceeds
the correlation length, $\xi_0$, of the order parameter fluctuations.
At the quantitative level, the instanton action $S$ can be
estimated as follows. To produce a nonzero DOS
at $E<E_g$, the optimal fluctuation of $\delta\Delta(\br)$
should have the magnitude of $-(E_g-E)$ and the spacial
extent of $L_E$, which immediately gives the estimate [S3]
\be
\label{S-estim}
  S_d
  \sim
  \frac{\Delta_0^2\xi^d}{f(0)}
  \left( \frac{E_g-E}{E_g} \right)^{2-d/4}
  .
\ee

To find the numerical coefficient in Eq.~(\ref{S-estim}),
one has to solve the instanton equation.
Measuring coordinates in terms of $L_E$ introduced in Eq.~(\ref{SM:L(E)}),
appropriately rescaling $\psi-\psi_0$, and replacing $\sinh\psi$
in the right-hand side of Eq.~(\ref{Usadel-psi}) by its value at $E_g$,
we rewrite Eq.~(\ref{Usadel-psi}) in a universal dimensionless form:
\be
  - \nabla^2 \phi + \phi - \phi^2 = h(\br) ,
\ee
where
\be
  \delta\Delta(\br)
  =
  -
  \frac{\Delta_0\Omega^2}{\rho\sinh\psi_g} \,
  h(\br/L_E) .
\ee

Minimization of the functional $\int h^2(\br) \, d^dr$
leads to the fourth order differential equation
for $\phi(\br)$:
\be
\label{Usadel2}
  ( -\nabla^2 + 1 - 2\phi )
  ( -\nabla^2\phi + \phi - \phi^2 ) = 0 .
\ee
The spherically symmetric optimal fluctuation solving (\ref{Usadel2})
in $d$ dimensions satisfies the second-order differential equation [S4]
\be
\label{inst-(d-2)}
  - \nabla^2_{d-2} \phi + \phi - \phi^2 = 0 ,
\ee
where $\nabla^2_{d-2}\equiv\partial^2/\partial r^2-(d-3)r^{-1}\partial/\partial r$
is the radial part of the Laplace operator in the $(d-2)$-dimensional space.
The instanton is characterized by the number
\be
  a_d
  = \int h^2 d^dr
  = 4 \int \frac{\phi'^2}{r^2} \, d^dr
  =
  \begin{cases}
    48\pi/5, & d=3, \\
    4.1, & d=2 ,
  \end{cases}
\ee
where $a_3$ follows from the exact solution
$\phi_3(r) = (3/2)\cosh^{-2}(r/2)$ [S3],
while $a_2$ is obtained by a numerical solution of Eq.~(\ref{inst-(d-2)}).

Returning to the dimensional variables, we get for the instanton action
in the limit $\eta\ll1$:
\be
  S_d
  =
  \frac{8a_d}{6^{d/4}}
  \frac{\Delta_0^2\xi^d}{f(0)}
  \left( \frac{E_g-E}{E_g} \right)^{2-d/4}
  .
\ee
In the 2D case,
\be
  S_2
  =
  \frac{4a_2}{\sqrt{6}}
  \frac{D\Delta_0}{f(0)}
  \left( \frac{E_g-E}{E_g} \right)^{3/2}
  ,
\ee
leading to Eqs.~(\ref{DOS-tail}) and (\ref{Gamma-tail}).

\section{IV. Role of a finite film thickness}

In films with a finite thickness $d\lesssim\xi_0$, there exists a contribution
to the depairing parameter $\eta$ in Eq.~(\ref{eta}) coming from
large wave vectors ($qd\gg1$), where diffusion is three-dimensional:
\be
\label{eta3D}
  \eta_\text{3D}
  =
  \frac{2}{\Delta}
  \int
  \frac{f(\bq)}{Dq^2}
  \frac{d^3\bq}{(2\pi)^3}
  =
  \frac{1}{\pi^2\Delta D}
  \int_{1/d}^{1/l} f(q) \, dq
\ee
[here $\Delta\equiv\Delta(T)$ is the temperature-dependent BCS order parameter].
In this region, Coulomb effects are weak and all complications related
with the energy dependence of $\lambda$ and $\Delta$ can be neglected.
Equation (\ref{Delta1-res}) is then replaced by a simpler expression
written for an arbitrary 3D wave vector:
\be
\label{Delta1-res-3D}
  \delta\Delta(T,\bq)
  =
  L_q \Bigl(\frac{T}{T_c}\Bigr) \,
  2\pi T
  \sum_{\eps>0}
  F_\text{dis}(\eps,\bq)
  ,
\ee
where $L_q(T/T_c)$ is the BCS fluctuation propagator at finite momentum:
\be
  L_q^{-1}(T/T_c)
  =
  2\pi T
  \sum_{\eps>0}
    \frac{\mathfrak{E}(\eps) Dq^2 + 2\Delta^2(T)}
      {\mathfrak{E}^2(\eps) [Dq^2 + 2\mathfrak{E}(\eps)]} .
\ee
In the limit $q\xi_0\gg1$, one recovers the known inverse logarithmic decay
of the fluctuation propagator \cite{SM:LO71}:
\be
\label{L-large-q}
  L_q(T/T_c)
  \approx
  \frac{1}{\ln(Dq^2/T_c)} ,
\qquad
  q\xi_0\gg1  .
\ee

With the help of Eq.~(\ref{<FF>})
the correlation function $f(\bq)=\corr{\delta\Delta\delta\Delta}_\bq$
can be written as
\be
  f(\bq) =
  L_q^2 \Bigl(\frac{T}{T_c}\Bigr) \,
  \left( \frac{T \Delta}{\nu_3} \right)^2
  \sum_{\eps_1,\eps_2>0}
  \int \frac{d^3\bk}{(2\pi)^3}
  \frac{\Pi_{\eps\eps'}(\bk+\bq/2) \Pi_{\eps\eps'}(\bk-\bq/2)}
  {\mathfrak{E}(\eps_1)\mathfrak{E}(\eps_2)} ,
\ee
where $\nu_3=\nu/d$ is the 3D DOS at the Fermi level.
Integrating over $\bk$ with the help of the Feynman's trick we get
\be
  f(\bq) =
  \frac{1}{8\pi}
  L_q^2 \Bigl(\frac{T}{T_c}\Bigr) \,
  \left( \frac{T \Delta}{\nu_3 D} \right)^2
  \sum_{\eps_1,\eps_2>0}
  \frac{1}
  {\mathfrak{E}(\eps_1)\mathfrak{E}(\eps_2)}
  \int_0^1
  \frac{dx}
  {[x(1-x)q^2+(\mathfrak{E}(\eps_1)+\mathfrak{E}(\eps_2))/D]^{1/2}} .
\ee

In the limit $q\xi_0\gg1$, only large $\eps_{1,2}\gg T_c$
with $\mathfrak{E}(\eps)\approx|\eps|$ are important.
Replacing summations by integrations and using Eq.~(\ref{L-large-q})
we find
\be
  f(\bq) =
  \frac{1}{8\pi^3}
  L_q^2 \Bigl(\frac{T}{T_c}\Bigr) \,
  \left( \frac{\Delta}{\nu_3 D} \right)^2
  \int_{T_c}^{Dq^2}
  \frac{d\eps_1}{\eps_1}
  \frac{d\eps_2}{\eps_2}
  \int_0^1
  \frac{dx}
  {[x(1-x)q^2+(\eps_1+\eps_2)/D]^{1/2}}
  \approx
  \frac{1}{8\pi^2q}
  \left( \frac{\Delta}{\nu_3 D} \right)^2 .
\ee

The obtained $1/q$ dependence of the correlation function $f(q)$
leads in Eq.~(\ref{eta3D}) to the logarithmic contribution
from the region $l\ll r\ll d$:
\be
\label{eta3D}
  \eta_\text{3D}
  =
  \frac{\Delta(T)}{8\pi^4 D}
  \left( \frac{1}{\nu_3 D} \right)^2
  \ln\frac dl ,
\ee
leading to Eq.~(\ref{eta3D-res}).

For even thicker films with $d\gg\xi_0$, the 2D contribution
(\ref{eta-2D}) is small and Eq.~(\ref{eta3D}) with $d$ replaced
by $\xi_0$ gives the leading contribution to the depairing rate.

\end{document}